\documentclass{article}
\usepackage[top=1in, bottom=1in, left=1in, right=1in]{geometry} 
\usepackage{setspace} 
\usepackage{amsfonts}
\usepackage{amsmath}
\usepackage{amsthm} 
\usepackage{amssymb} 
\usepackage{url}
\usepackage{bm}
\usepackage{xcolor}
\usepackage{mathtools} 
\usepackage{booktabs}
\usepackage{enumitem} 

\usepackage{graphicx}
\usepackage{subcaption}
\usepackage[colorlinks,pagebackref=true]{hyperref}

\DeclareMathOperator{\spn}{span}
\DeclareMathOperator{\Ima}{Im}

\DeclareMathOperator{\rank}{rank}

\DeclareMathOperator*{\argmin}{arg\,min}
\DeclareMathOperator{\rec}{rec}

\newcommand{\e}{{\rm e}}

\newcommand{\R}{{\mathbb R}}

\newcommand{\Acal}{{\mathcal A}}

\newcommand{\Ccal}{{\mathcal C}}

\newcommand{\Gcal}{{\mathcal G}}

\newcommand{\Ncal}{{\mathcal N}}

\newcommand{\Pcal}{{\mathcal P}}
\newcommand{\Qcal}{{\mathcal Q}}

\newtheorem{proposition}{Proposition}[section]
\newtheorem{lemma}[proposition]{Lemma}
\newtheorem{theorem}[proposition]{Theorem}
\newtheorem{definition}[proposition]{Definition}
\newtheorem{corollary}[proposition]{Corollary}
\newtheorem{remark}[proposition]{Remark}

\newtheorem{assumption}[proposition]{Assumption}

\newtheorem{exampleemph}[proposition]{Example}   
\newenvironment{example}{\begin{exampleemph}\begin{upshape}}{\end{upshape}\end{exampleemph}} 

\newcommand\tcapfig[1]{\captionsetup{position=top, font=normalsize, labelfont=bf, textfont=normalfont, justification=centering, margin=0mm, aboveskip=2mm, belowskip=0mm, labelsep=colon, singlelinecheck=false}\caption{#1}}
\newcommand\bnotefig[1]{\captionsetup{position=bottom, font=footnotesize,  textfont=normalfont, margin=1mm, skip=2mm, justification=justified, singlelinecheck=false}\caption*{#1}}
\begin{document}

\title{Fixed-Income Pricing and the Replication of Liabilities\footnote{Part of
this paper was written while the author was a guest of the FinsureTech Hub,
Department of Mathematics, ETH Zurich. The author thanks an anonymous referee,
whose questions on the tightness of the super-replication bounds and on the
smoothness of fitted discount curves prompted Theorem~\ref{thm_cs} and
Section~\ref{sec_num}, and Rose Wang for excellent research assistance in
constructing the Treasury data set used there.}}
\author{ Damir Filipovic\footnote{EPFL, Swiss Finance Institute, 1015 Lausanne, Switzerland. Email: damir.filipovic@epfl.ch}}
\date{\today}
\maketitle

\begin{abstract}
This paper develops a model-free framework for static fixed-income pricing and
the replication of liability cash flows. The absence of static arbitrage across
a universe of fixed-income instruments is equivalent to the existence of a
strictly positive discount curve reproducing all observed prices. Linear
programming duality then identifies the least-cost super-replication price with
the largest value that any admissible discount curve assigns to the liability,
so that the resulting bounds are attained and cannot be improved. Complementary
slackness confines over-replication to dates that the optimal discount vector
prices at zero, and a least-cost portfolio matches the liability exactly at no
fewer dates than the rank of the cash-flow matrix. We also obtain generic
uniqueness of that portfolio, an interpolation between quadratic hedging and
super-replication, and a static treatment of swap--repo strategies. On US
Treasury cross-sections the observed prices violate the law of one price, so
that a discount curve must be estimated rather than bootstrapped; the least-cost
portfolio then matches an annuity liability at almost every cash-flow date.
\end{abstract}



\section{Introduction}\label{sec_intro}

The valuation and replication of insurance liabilities rely fundamentally on the
structure of fixed-income markets. Regulatory frameworks such as Solvency~II and
the Swiss Solvency Test require insurers to discount expected liability
cash flows using market-consistent yield curves and to assess hedging strategies
based on portfolios of bonds, swaps, and repo transactions.\footnote{For Solvency II see \url{https://www.eiopa.europa.eu/browse/regulation-and-policy/solvency-ii_en}. For the Swiss Solvency Test see \url{https://www.finma.ch/en/supervision/insurers/cross-sectoral-tools/swiss-solvency-test-sst}.} Despite the
operational importance of these practices, a basic structural question remains:
under what conditions do the observed prices of fixed-income instruments define
an arbitrage-free pricing system that admits a common discount curve and
supports a consistent valuation and replication of expected liability cash
flows?

This paper develops a general and model-free framework for fixed-income pricing
based solely on static arbitrage. We show that the absence of (strict) arbitrage
across a universe of fixed-income instruments is equivalent to the existence of
a strictly positive (nonnegative) discount curve that reproduces all observed
market prices. This result constitutes a fundamental theorem of fixed-income pricing and
provides a theoretical foundation for discount-curve construction in insurance
asset--liability management, as studied, for example, in
\cite{Filipovic2022,CamenzindFilipovic2024}. To the best of our knowledge, this
paper provides the first systematic treatment of static arbitrage in fixed-income
price systems formulated directly in terms of dated cash flows and observed
market prices, with an explicit focus on liability replication.

We then turn to the replication of expected insurance liability cash flows.
Perfect replication typically fails when liability and asset cash-flow dates and
cash flows do not align. This motivates a super-replication approach, for which
we establish feasibility and existence of a least-cost super-replicating
portfolio. Linear programming duality then shows that the least-cost
super-replication price equals the largest value assigned to the liability cash
flows by any nonnegative discount curve consistent with observed prices, and
that the corresponding sub-replication price is given by the smallest such
value. The resulting interval is exactly the set of liability values that can be
adjoined to the market without creating a strict arbitrage; its endpoints are
attained and therefore cannot be improved, and the interval degenerates to a
single point precisely when the liability is exactly replicable. Complementary
slackness sharpens this further: strict over-replication can occur only at dates
to which the optimal discount vector assigns zero value, so that the excess cash
flows of a least-cost super-replicating portfolio carry no value, and some
least-cost portfolio matches the liability exactly at no fewer dates than the
rank of the cash-flow matrix. We also show that the least-cost super-replicating
portfolio is unique for Lebesgue-almost every arbitrage-free price vector, which
settles a natural question left open by the linearity of the objective, and that
a family of asymmetric quadratic hedging criteria interpolates between quadratic
hedging and least-cost super-replication. A key insight for implementation is
that swap--repo strategies naturally produce fixed-income cash-flow profiles,
allowing swaps, repos, and floating-rate notes to be treated within the same
static framework as coupon bonds.

Our first main result characterizes the existence of a discount curve consistent
with observed prices, and thereby the entire set $\Gcal$ of such curves. On a
cross-section of coupon bonds with interlocking cash flows the cash-flow matrix
is far from having full rank, with two consequences. First, when $\Gcal$ is
nonempty it is of large dimension, and individual elements of it may imply
irregular forward-rate shapes. Second, and more starkly, $\Gcal$ may be empty:
Section~\ref{sec_num} finds that it is, on every US Treasury cross-section
examined, so that prices must be treated as observed with error and a discount
curve estimated. Selecting a particular, smooth element of $\Gcal$, or an approximating
curve when prices are observed with noise, is the subject of a large literature
on term-structure estimation
\cite{nel_sie_87,sve_94,gur_sac_wri_07,filipovic:exact_term_structure,liu_wu_21,Filipovic2022,CamenzindFilipovic2024};
it is also what motivates the practice, common among government bond dealers, of
working with curves that price bonds on a least-squares rather than an exact
basis, the residual pricing errors themselves carrying information about market
liquidity \cite{hu_pan_wang_13}. This selection problem is logically distinct
from the existence question studied here. As we show in
Remark~\ref{rem_selection}, it also leaves the replication bounds of
Section~\ref{sec_repli} unaffected, since these depend on the market only
through the set $\Gcal$ as a whole and not on any particular curve drawn from
it.

The results presented here are consistent with standard fixed-income valuation
and replication practices and clarify the conditions under which current
regulatory discounting and liability valuation methodologies are well founded,
as well as their structural limitations. They also point toward several avenues
for further research, including numerical methods for liability hedging and
implications for the construction of regulatory discount curves.

The theoretical foundations of arbitrage-free pricing are well established in
the discrete-time and static asset pricing literature. Seminal contributions by \cite{har_kre_79} and \cite{dal_mor_wil_90} establish the
equivalence between arbitrage-free price systems and the existence of linear
pricing functionals, or state-price vectors; see \cite{coc_01} and
\cite{foe_sch_16} for surveys and further references. These results apply to general financial markets and are not tailored
to fixed-income price systems described directly in terms of dated cash flows.
To the best of our knowledge, a systematic treatment of static arbitrage in
fixed-income markets formulated explicitly at the level of cash-flow matrices
and observed prices, and linked to the replication of expected liability
cash flows, has not appeared in the literature.

The remainder of the paper is organized as follows.
Section~\ref{sec_FTFIP} develops the fundamental theorem of fixed-income pricing
and establishes the equivalence between arbitrage-free prices and the existence
of a discount curve.
Section~\ref{sec_repli} applies this framework to the replication and
super-replication of liabilities: it derives conditions under which least-cost
super-replication is feasible, establishes the duality and tightness results
described above, addresses uniqueness, and develops asymmetric quadratic hedging
as a smooth relaxation.
Section~\ref{sec_exam} illustrates the theory through canonical fixed-income
instruments, including coupon bonds, swaps, and repo-based strategies.
Section~\ref{sec_num} implements the framework on cross-sections of US
Treasury securities, documenting the failure of the law of one price at observed
prices, the estimation step this makes necessary, and the tightness of the
replication bounds for an annuity liability.
Section~\ref{sec_conc} concludes with an outlook on open questions and directions
for further research.
All proofs are collected in the Appendix.

\section{A Fundamental Theorem of Fixed-Income Pricing}
\label{sec_FTFIP}

We consider a universe of $M$ fixed-income instruments, indexed by
$i = 1,\dots,M$, with prices $P_i$ and deterministic cash flows
$C_{i,1},\dots,C_{i,N}$ occurring at a common set of dates
$0 < x_1 < \cdots < x_N$.%
\footnote{If instrument $i$ has no cash flow at date $x_j$, we set $C_{i,j} = 0$.}
We collect the prices in the vector $P \in \mathbb{R}^M$ and the cash flows in the
$M \times N$ cash-flow matrix $C$. We write $C_{i,:}$ for the $i$th row of $C$,
the cash-flow profile of instrument~$i$, and $C_{:,j}$ for the $j$th column, the
cash flows of all instruments at date~$x_j$. We also denote by
$\bm{x} \coloneq [x_1,\dots,x_N]^{\top}$ the vector of cash-flow dates, and for
any function $g \colon [0,\infty) \to \mathbb{R}$ we write
$g(\bm{x}) \coloneq [g(x_1),\dots,g(x_N)]^{\top}$.

We assume a frictionless market in which these instruments can be traded
freely. A portfolio is represented by a vector $q \in \mathbb{R}^M$, where $q_i$
denotes the position in instrument~$i$. Such a portfolio has price
$q^{\top} P$ and generates the cash flows $q^{\top} C$.

A central question, which has received surprisingly little attention in the
literature, is whether the observed price system is \emph{internally
consistent}, in the sense that it does not permit static arbitrage across the
fixed-income instruments. To address this question, we formalize the standard
notions of static arbitrage in this setting. We begin with the most basic
consistency requirement, namely that prices depend linearly on cash flows.

\begin{theorem}[Law of one price]\label{thm_LoP}
The following statements are equivalent:
\begin{enumerate}[label=\textnormal{(L\arabic*)}]
    \item\label{LoP1} The \emph{law of one price} holds: every portfolio $q\in\R^M$ that generates zero cash flows, $q^\top C=0$, has zero price, $q^\top P=0$. 
    \item\label{LoP2} There exists a real-valued discount curve $g:[0,\infty)\to\R $ with $g(0)=1$ and such that $P = C\, g(\bm{x})$.
\end{enumerate}    
\end{theorem}

This basic consistency property is strengthened by the following notions of
arbitrage.

\begin{definition}\label{defNA}
A portfolio $q \in \mathbb{R}^M$ is called
\begin{enumerate}
\item a \emph{strict arbitrage} if it has strictly negative price,
      $q^{\top}P < 0$, and generates nonnegative cash flows,
      $q^{\top}C \ge 0$;

\item an \emph{arbitrage} if it has nonpositive price, $q^{\top}P \le 0$, and
      generates nonnegative cash flows, $q^{\top}C \ge 0$, with at least one
      strictly positive component, that is, $(q^{\top}C)_j > 0$ for some
      $j \le N$.
\end{enumerate}
\end{definition}

We now derive the fundamental characterization of arbitrage in fixed-income
pricing. To this end, define
\[
\mathcal{C} \coloneq \{\, C^{\top} q : q^{\top}P < 0 \,\},
\qquad
\overline{\mathcal{C}} \coloneq \{\, C^{\top} q : q^{\top}P \le 0 \,\},
\]
the sets of cash-flow vectors generated by portfolios with strictly negative and
nonpositive prices, respectively.
By construction, $\mathcal{C}$ is an open convex cone in $\mathbb{R}^N$, 
$\overline{\mathcal{C}}$ is a closed convex cone, and indeed 
$\overline{\mathcal{C}}$ is the closure of $\Ccal$.

The next two theorems characterize, first, the absence of strict arbitrage
and, second, the absence of arbitrage, together with their implications for
fixed-income pricing.

\begin{theorem}[Absence of strict arbitrage]\label{thm:wNA}
The following statements are equivalent:
\begin{enumerate}[label=\textnormal{(W\arabic*)}]
    \item\label{wNA1} There is no strict arbitrage.
    \item\label{wNA2} $\mathcal{C} \cap \mathbb{R}^N_{+} = \emptyset$.
    \item\label{wNA3} There exists a nonnegative discount curve $g \ge 0$ with $g(0)=1$ and such that $P = C\, g(\bm{x})$.
\end{enumerate}

\noindent
Moreover, each of the conditions \ref{wNA1}--\ref{wNA3} implies the law of one price \ref{LoP1}--\ref{LoP2} in Theorem~\ref{thm_LoP}.
\end{theorem}

The characterization of the absence of arbitrage requires one additional, mild
condition on the traded universe.

\begin{assumption}[Nondegeneracy]\label{ass_ND}
There exists a portfolio $q\in\R^M$ with $q^\top C\ge 0$ and $q^\top C\neq 0$.
\end{assumption}

Assumption~\ref{ass_ND} holds as soon as one traded instrument has nonnegative,
not identically zero cash flows, $C_{i,:}\ge 0$ and $C_{i,:}\neq 0$; any bond,
bill, or long repo position will do. It is therefore innocuous in any
application we have in mind. It is not, however, implied by the remaining
conditions, and it cannot be dispensed with, as Example~\ref{ex_ND} below shows.

\begin{theorem}[Absence of arbitrage]\label{thm:NA}
The following statements are equivalent, provided Assumption~\ref{ass_ND} holds:
\begin{enumerate}[label=\textnormal{(S\arabic*)}]
    \item\label{sNA1} There is no arbitrage.
    \item\label{sNA2} $\overline{\mathcal{C}} \cap \mathbb{R}^N_{+} = \{0\}$.
    \item\label{sNA3} There exists a strictly positive discount curve $g > 0$ with $g(0)=1$ and such that $P = C\, g(\bm{x})$.
\end{enumerate}

\noindent
Moreover, each of the conditions \ref{sNA1}--\ref{sNA3} then implies each
of the weak conditions \ref{wNA1}--\ref{wNA3} in
Theorem~\ref{thm:wNA}, and hence also the law of one price
\ref{LoP1}--\ref{LoP2}. Without Assumption~\ref{ass_ND}, the equivalence
\ref{sNA1}$\Leftrightarrow$\ref{sNA2} and the implication
\ref{sNA3}$\Rightarrow$\ref{sNA2} remain valid.
\end{theorem}

Here is an example illustrating that no arbitrage is strictly stronger than no strict arbitrage.

\begin{example}
Consider $M = N = 2$ fixed-income instruments with prices $P_1 = 1$ and $P_2 = c$,
and cash-flow rows $C_{1,:} = [1,0]$ and $C_{2,:} = [c,1]$, occurring at dates
$0 < x_1 < x_2$, where $c > 0$ is a fixed coupon rate.  
The pricing equations imply that any discount curve must satisfy $g(x_1) = 1$ and $g(x_2) = 0$.
Thus the discount curve is nonnegative but not strictly positive.  
Consequently, there is no strict arbitrage, but arbitrage opportunities exist.

Indeed, the portfolio $q = [-c,\,1]^{\top}$ has price
$q^{\top}P =   0$ and generates the nonnegative, nonzero cash-flow
vector $q^{\top}C =  [0,1]$.  
Hence $q$ is an arbitrage but not a strict arbitrage.
\end{example}

The next example shows that Assumption~\ref{ass_ND} is indispensable in
Theorem~\ref{thm:NA}.

\begin{example}\label{ex_ND}
Consider $M=N=2$ instruments with cash-flow rows $C_{1,:} = [1,-1]$ and
$C_{2,:} = [2,-2]$ and prices $P = [1,1]^{\top}$. Every portfolio generates the
cash flow $q^{\top} C = (q_1+2q_2)\,[1,-1]$, which is never nonnegative and
nonzero. Hence there is no arbitrage, \ref{sNA1} holds, and
Assumption~\ref{ass_ND} fails. On the other hand
$\Ima(C) = \{ a\,[1,2]^{\top} : a\in\R\}$ does not contain $P$, so there exists
no discount curve at all with $P = C\,g(\bm x)$, and \ref{sNA3} fails. Indeed,
the law of one price fails as well: the portfolio $q = [2,-1]^{\top}$ has zero
cash flows and price $1$, and $-q$ is a strict arbitrage.

Economically, the two instruments have proportional net cash-flow profiles---two
receiver swaps with proportional net legs, say---quoted at inconsistent prices.
The resulting free lunch is a strict arbitrage, but not an arbitrage in the
sense of Definition~\ref{defNA}(ii), because no portfolio in this universe
produces a nonnegative nonzero cash flow.
\end{example}

It is convenient to collect the discount factors consistent with the observed
prices in the set
\begin{equation}\label{eqGcal}
    \Gcal \coloneq \{\, v\in\R^N_{+} \;:\; C v = P \,\},
\end{equation}
a closed convex polyhedron. By Theorem~\ref{thm:wNA}, $\Gcal\neq\emptyset$ if
and only if there is no strict arbitrage, and $v\in\Gcal$ if and only if
$v = g(\bm x)$ for some nonnegative discount curve $g$ as in \ref{wNA3}. By
Theorem~\ref{thm:NA}, $\Gcal$ contains a strictly positive vector if and only if
there is no arbitrage.

\begin{remark}[Multiplicity and curve selection]\label{rem_mult}
The set $\Gcal$ is a singleton only in the exceptional case
$\ker(C)\cap(\Gcal-\Gcal) = \{0\}$. The relevant comparison here is not between
the number $M$ of instruments and the number $N$ of cash-flow dates---on a
cross-section of government bonds the former typically exceeds the
latter---but between $\rank(C)$ and $N$. Coupon dates cluster on a common
grid, and the grid thins with maturity, so that $\rank(C) < N$ with the
deficiency concentrated at long maturities, where few instruments pay and their
restricted cash-flow profiles are nearly proportional. The pricing equations
then pin down the discount curve only on an affine slice of $\R^N_{+}$, and not
at all between the cash-flow dates. Any procedure that returns a single
curve---bootstrapping, spline fitting, kernel ridge regression---therefore
embeds a selection criterion in addition to the pricing constraints.

The indeterminacy is uneven across dates. Where two cash flows of different
instruments fall close to one another, the traded prices pin down their combined
value but not the split between them, and a curve reproducing the prices exactly
is free in that direction. Exact-fit curves are correspondingly known to imply
irregular forward-rate shapes, which is one reason they are commonly replaced in
practice by least-squares fits
\cite{nel_sie_87,sve_94,gur_sac_wri_07,filipovic:exact_term_structure,liu_wu_21,Filipovic2022,CamenzindFilipovic2024,hu_pan_wang_13}.
Section~\ref{sec_num} quantifies the indeterminacy on a US Treasury
cross-section, where the prices determine 223 of 249 discount factors exactly
and leave 26 free.

Theorem~\ref{thm:NA} is silent on this selection, by design: it characterizes
$\Gcal\neq\emptyset$, not which of its elements should be used for
interpolation. Remark~\ref{rem_selection} below shows that the replication
bounds of Section~\ref{sec_repli} do not require such a selection at all.
\end{remark}

\section{Liability Replication and Super-Replication}\label{sec_repli}

We consider a liability portfolio that generates expected cash flows
$Z = [Z_1,\dots,Z_N]$ at the dates $x_j$.%
\footnote{Expected cash flows are understood as
expectations taken under the relevant risk-neutral forward measures. Under
standard assumptions in insurance liability modeling, these expectations
coincide with expectations under the real-world measure.}\footnote{As for the fixed-income instruments, $Z_j = 0$ indicates that no expected
liability cash flow occurs at date $x_j$.} Our aim is to replicate these liability cash flows using a portfolio in the 
fixed-income instruments.  
In the strict sense, this means finding a portfolio $q \in \mathbb{R}^M$ such that
\begin{equation}\label{eqreplistrict}
    q^{\top} C = Z.
\end{equation}
A portfolio $q$ satisfying \eqref{eqreplistrict} exists if and only if 
$Z^{\top} \in \operatorname{Im}(C^{\top})$, or equivalently, 
$\ker(C) \subseteq \ker(Z)$.  
In practice, this requirement may be too strong and need not hold.  
We therefore relax perfect replication to \emph{super-replication} and define the 
super-replication set
\[
    \mathcal{Q} \coloneqq \{\, q \in \mathbb{R}^M : q^{\top} C \ge Z \,\}.
\]
It is immediate that $\mathcal{Q}=\cap_{j\le N}\{\, q \in \mathbb{R}^M : q^{\top} C_{:,j} \ge Z_j \,\}  $ is a closed convex polyhedron in $\mathbb{R}^{M}$,
and every portfolio satisfying \eqref{eqreplistrict} belongs to $\mathcal{Q}$. 

The following result characterizes the feasibility of super-replication.
It shows that the super-replication set $\mathcal{Q}$ is nonempty if and only if
there exists no nonnegative discount vector that assigns zero value to all
attainable cash flows while assigning strictly positive value to the liability
cash flows~$Z$.

\begin{theorem}[Well-posedness]\label{thm_WP}
The following statements are equivalent:
\begin{enumerate}[label=\textnormal{(F\arabic*)}]
    \item\label{F1} Super-replication is feasible, that is, $\mathcal{Q}\neq\emptyset$.
    \item\label{F2} There exists no vector $v\in\mathbb{R}^N_{+}$ such that
    $C v = 0$ and $Z v > 0$.
\end{enumerate}

\noindent
A sufficient condition for \ref{F1}--\ref{F2} to hold is the existence of a
fixed-income instrument $i$ with nonnegative cash flows whose effective
cash-flow dates coincide with those of the liabilities, in the sense that
$Z_j \neq 0$ if and only if $C_{i,j} > 0$ for all $j \le N$.
\end{theorem}

Theorem~\ref{thm_WP} has a simple necessary condition as an immediate
consequence. Let $Z\ge 0$ and suppose $Z_j>0$ at a date $x_j$ at which no traded
instrument pays, so that $C_{:,j}=0$. Then $v=e_j$ satisfies $Cv=0$ and
$Zv=Z_j>0$, so that \ref{F2} fails and $\Qcal=\emptyset$. The effective cash-flow dates of the liabilities must
therefore be contained in those of the traded instruments.

This is restrictive in applications, since insurance liabilities are paid
monthly or on policy dates that need not coincide with bond coupon dates. When
alignment fails one may work instead with a \emph{proxy} cash-flow matrix,
obtained by aggregating intermediate payments between liability dates. This
proxy has a natural operational interpretation if intermediate cash proceeds can
be held in a cash buffer, or, under no arbitrage, reinvested at
forward-equivalent rates when the corresponding instruments are available.

\begin{remark}[Cash-flow aggregation]\label{rem_agg}
Let $x_{j_1} < \cdots < x_{j_n}$ denote the liability payment dates, that is,
$Z_{j_k} \neq 0$.  For each fixed-income instrument $i$, define a modified
cash-flow matrix $\tilde C$ by aggregating payments between liability payment dates: with
$j_0 \coloneq 0$ and for $k=1,\dots,n$,
\[
\tilde C_{i,j} \coloneq
\begin{cases}
0, & j_{k-1} < j < j_k,\\
\sum_{j_{k-1} < s \le j_k} C_{i,s}, & j = j_k,\\
C_{i,j}, & j > j_n.
\end{cases}
\]
Operationally, this corresponds to assuming that intermediate cash flows can be
held in a cash buffer until the next liability payment date.

Alternatively, under no arbitrage~\ref{sNA1}, intermediate cash flows may be accumulated at
forward-equivalent rates using a strictly positive discount curve $g>0$ by
setting
\[
\tilde C_{i,j_k}
= \sum_{j_{k-1} < s \le j_k} \frac{g(x_s)}{g(x_{j_k})}\, C_{i,s}.
\]
In this case $g$ is also a discount curve for the modified cash-flow matrix, 
$P = C\,g(\bm{x}) = \tilde C\,g(\bm{x})$.  This construction presumes the
availability of the corresponding forward rate agreements.
\end{remark}

\subsection{Least-cost super-replication and duality}

Throughout we use the conventions $\inf\emptyset = +\infty$ and
$\sup\emptyset = -\infty$, since the feasible sets below may be empty.

We now seek a super-replication portfolio of minimal initial cost, that is, a
solution to the linear program
\begin{equation}\label{suprepeq}
    \pi^{+}(Z) \coloneq \inf_{q \in \mathcal{Q}} q^{\top} P ,
\end{equation}
and, symmetrically, the largest amount that can be raised today against the
liability without any risk of over-delivering,
\begin{equation}\label{subrepeq}
    \pi^{-}(Z) \coloneq \sup\{\, q^{\top} P \;:\; q^{\top} C \le Z \,\}.
\end{equation}
The following duality theorem is the central result of this section. It
identifies the least-cost super-replication price with the largest value that
any admissible discount curve assigns to the liability cash flows, and it
delivers existence of a least-cost super-replicating portfolio under the weakest
of our no-arbitrage conditions.

\begin{theorem}[Super-replication duality]\label{thm_dual}
Assume there is no strict arbitrage~\ref{wNA1}. Then
\begin{equation}\label{eqdual}
    \pi^{+}(Z) \;=\; \sup_{v\in\Gcal} Z v \;\in\; (-\infty,+\infty] ,
    \qquad
    \pi^{-}(Z) \;=\; \inf_{v\in\Gcal} Z v \;\in\; [-\infty,+\infty) ,
\end{equation}
and $\pi^{-}(Z)\le\pi^{+}(Z)$. Moreover, $\pi^{+}(Z)<\infty$ if and only if
super-replication is feasible~\ref{F1}, in which case the infimum in
\eqref{suprepeq} and the supremum in \eqref{eqdual} are both attained;
analogously for $\pi^{-}(Z)$ and the feasibility of \eqref{subrepeq}.
\end{theorem}

The two hypotheses of Theorem~\ref{thm_dual} are precisely dual and primal
feasibility: \ref{wNA1} states that $\Gcal\neq\emptyset$, and \ref{F1} states
that $\Qcal\neq\emptyset$. No further regularity is needed; in particular,
neither the absence of arbitrage nor the injectivity condition
$\ker(C^{\top})=\{0\}$ is required for existence.

The economic content of \eqref{eqdual} is that the super-replication price is
the largest value that the liability can be assigned by a discount curve
consistent with the observed market prices. The bounds are therefore not merely
valid, but attained, in the following sense.

\begin{corollary}[The bounds are the best possible]\label{cor_interval}
Assume \ref{wNA1}. Consider the market enlarged by the liability, regarded as an
instrument with cash flows $Z$ and price $p$. Then the enlarged market admits no
strict arbitrage if and only if $p\in\{\,Zv : v\in\Gcal\,\}$, and this set is the
interval with endpoints $\pi^{-}(Z)$ and $\pi^{+}(Z)$, each endpoint belonging
to it whenever finite. If \eqref{suprepeq} and \eqref{subrepeq} are both
feasible, it is the compact interval $[\pi^{-}(Z),\pi^{+}(Z)]$.
\end{corollary}

Consequently, no super-replicating portfolio costs less than $\pi^{+}(Z)$, and
$\pi^{+}(Z)$ cannot be lowered without admitting a strict arbitrage. The width
$\pi^{+}(Z)-\pi^{-}(Z)$ of the interval is a property of the traded universe
relative to the liability, and not slack in the method: it vanishes exactly when
the liability is attainable.

\begin{corollary}[Collapse of the interval]\label{cor_collapse}
Let Assumption~\ref{ass_ND} hold, assume there is no arbitrage~\ref{sNA1} and
that \eqref{suprepeq} and \eqref{subrepeq} are feasible. Then
$\pi^{-}(Z)=\pi^{+}(Z)$ if and only if $Z^{\top}\in\Ima(C^{\top})$, that is, if
and only if the liability is exactly replicable in the sense of
\eqref{eqreplistrict}. In that case the common value equals $q^{\top}P$ for any
$q$ with $q^{\top}C = Z$.
\end{corollary}

For nonnegative liability cash flows, $Z\ge 0$, which is the relevant case in
insurance applications, the feasibility hypothesis on \eqref{subrepeq} in
Corollary~\ref{cor_collapse} is automatically satisfied, since $q=0$ is then
admissible in \eqref{subrepeq}; the hypothesis is stated explicitly only because
$Z$ is allowed to take negative entries throughout.

The next result concerns the tightness of the optimal portfolio itself, that is,
the extent to which the super-replicating cash flows $q^{\top}C$ hug the
liability cash flows~$Z$.

\begin{theorem}[Structure of least-cost super-replication]\label{thm_cs}
Assume \ref{wNA1} and \ref{F1}. Let $q$ be a solution of \eqref{suprepeq} and
let $v\in\Gcal$ be a solution of the dual problem in \eqref{eqdual}. Then:
\begin{enumerate}
\item\label{cs1} complementary slackness holds,
      $v_j\,\bigl[(q^{\top}C)_j - Z_j\bigr] = 0$ for all $j\le N$; equivalently,
      strict over-replication $(q^{\top}C)_j > Z_j$ can occur only at dates
      $x_j$ to which the optimal discount vector assigns zero value, $v_j = 0$;
\item\label{cs2} consequently $q^{\top}P = \sum_{j\le N} Z_j v_j$: valued at the
      optimal discount vector, the excess cash flows $q^{\top}C - Z$ cost
      nothing;
\item\label{cs3} if some dual solution satisfies $v>0$, then $q^{\top}C = Z$,
      that is, least-cost super-replication is exact;
\item\label{cs4} if in addition there is no arbitrage~\ref{sNA1}, then
      \eqref{suprepeq} admits a solution $q$ at which the active columns
      $\{C_{:,j} : (q^{\top}C)_j = Z_j\}$ span $\Ima(C)$. At any such solution
      the liability is matched exactly at no fewer than $\rank(C)$ dates, and
      strict over-replication occurs at no more than $N-\rank(C)$ dates. If
      moreover $\ker(C^{\top})=\{0\}$, so that $\rank(C)=M$, then $\Qcal$ is
      pointed and any such $q$ may be taken to be a vertex of $\Qcal$.
\end{enumerate}
\end{theorem}

Parts \ref{cs1}--\ref{cs2} say that a least-cost super-replicating portfolio
never pays for its own excess: over-delivery is confined to dates that the
extremal admissible discount vector prices at zero. Part~\ref{cs4} is the
complementary combinatorial statement: at least as many dates are matched
exactly as the rank of the cash-flow matrix, so that the scope for a loose,
uninformative super-replication is limited by the richness of the traded
universe. Redundant instruments do not help here, and the bound is stated in
terms of $\rank(C)$ rather than $M$ for that reason; the two coincide precisely
when $\ker(C^{\top})=\{0\}$.

\begin{remark}[Curve selection is immaterial for the bounds]\label{rem_selection}
The prices $\pi^{\pm}(Z)$ and the optimal portfolios depend on the market only
through the set $\Gcal$ of \eqref{eqGcal}, and not on the choice of a particular
discount curve within $\Gcal$. The irregularity of exactly fitted forward rates
discussed in Remark~\ref{rem_mult} therefore does not propagate into the
replication bounds: no element of $\Gcal$ needs to be singled out, interpolated,
or smoothed in order to compute $\pi^{\pm}(Z)$ or the corresponding portfolios. Smoothness
remains essential for the separate task of valuing cash flows at dates that are
not spanned by traded instruments, where a selection within $\Gcal$ is
unavoidable.
\end{remark}

\subsection{Uniqueness}

Uniqueness in \eqref{suprepeq} may fail: if two instruments have identical cash
flows and identical prices, any optimal position can be shifted between them
without affecting cost or cash flows. Such ties are, however, non-generic.

\begin{proposition}[Generic uniqueness]\label{prop_uniq}
Let Assumption~\ref{ass_ND} hold, fix $C$ and $Z$ with $\Qcal\neq\emptyset$ and
$\ker(C^{\top}) = \{0\}$, and let $\Pcal_{\rm na}\subseteq\R^{M}$ denote the set
of arbitrage-free price vectors, that is, those satisfying~\ref{sNA1}. Then
there exists a finite union $\Ncal$ of hyperplanes in $\R^{M}$, depending on $C$
and $Z$ only, such that \eqref{suprepeq} has a unique solution for every
$P\in\Pcal_{\rm na}\setminus\Ncal$. In particular, the least-cost
super-replicating portfolio is unique for Lebesgue-almost every arbitrage-free
price vector.
\end{proposition}

A canonical selection on the exceptional set $\Ncal$ is provided by
Theorem~\ref{thm_limit} below, which singles out the least-cost
super-replicating portfolio of minimal norm.

\subsection{Asymmetric quadratic hedging}\label{sec_asym}

Super-replication and quadratic hedging are the two extreme points of a single
family of criteria, indexed by the weight placed on shortfall relative to cost.
For $\kappa>0$ and $\lambda>0$ define
\begin{equation}\label{eqJ}
    J_{\kappa,\lambda}(q) \coloneq q^{\top} P
      + \kappa \sum_{j\le N}
        \Bigl(\bigl(Z_j - (q^{\top}C)_j\bigr)_{+}\Bigr)^{2}
      + \lambda \|q\|_2^{2} ,
\end{equation}
where $u_{+} = \max\{u,0\}$. The first term is the initial cost of the
portfolio, the second penalizes shortfall only, leaving over-replication
unpenalized, and the third is a ridge term controlling the size of the
positions. Other convex asymmetric losses, such as smoothed hinge penalties, can
be substituted for the squared shortfall without affecting the results below.

\begin{proposition}\label{prop_qexists}
For all $\kappa,\lambda>0$ the function $J_{\kappa,\lambda}$ is continuously
differentiable, strictly convex and coercive. It therefore admits a unique
minimizer $q_{\kappa,\lambda}\in\R^{M}$, characterized by the first-order
condition
\begin{equation}\label{eqFOC}
    C\, v_{\kappa,\lambda} = P + 2\lambda\, q_{\kappa,\lambda},
    \qquad
    v_{\kappa,\lambda} \coloneq 2\kappa
      \Bigl(\bigl(Z - q_{\kappa,\lambda}^{\top}C\bigr)_{+}\Bigr)^{\!\top}
      \;\in\; \R^{N}_{+} .
\end{equation}
\end{proposition}

The vector $v_{\kappa,\lambda}$ of scaled shortfalls thus plays the role of a
discount vector: it is nonnegative, it reprices the traded instruments up to the
error $2\lambda q_{\kappa,\lambda}$, and it vanishes at every date at which the
portfolio over-delivers. Letting $\lambda\downarrow 0$ and $\kappa\to\infty$
turns \eqref{eqFOC} into dual feasibility $v\in\Gcal$ together with the
complementary slackness of Theorem~\ref{thm_cs}\ref{cs1}. The following theorem
makes this precise at the level of the portfolios.

\begin{theorem}[Interpolation]\label{thm_limit}
Let Assumption~\ref{ass_ND} hold and assume \ref{sNA1}, \ref{F1} and
$\ker(C^{\top}) = \{0\}$. Then:
\begin{enumerate}
\item for every fixed $\lambda>0$, $q_{\kappa,\lambda}\to q_{\lambda}$ as
      $\kappa\to\infty$, where
      $q_{\lambda} \coloneq \argmin_{q\in\Qcal}\{q^{\top}P + \lambda\|q\|_2^2\}$
      is well defined;
\item $q_{\lambda}\to q^{\star}$ as $\lambda\downarrow 0$, where $q^{\star}$ is
      the minimal-norm element of the solution set of \eqref{suprepeq}.
\end{enumerate}
\end{theorem}

The asymmetric quadratic hedge therefore interpolates between smooth quadratic
hedging and least-cost super-replication, and the iterated limit selects a
distinguished least-cost super-replicating portfolio irrespective of the
cardinality of the solution set of \eqref{suprepeq}.

\begin{remark}[Symmetric quadratic hedging]
Penalizing over- and under-replication symmetrically, as in
$\|Z - q^{\top}C\|_2^2 + \lambda\|q\|_2^2$, also yields a unique solution, but
provides no control over the direction of the replication error, which is the
relevant asymmetry in liability-driven investment. The family \eqref{eqJ}
retains uniqueness and smoothness while restoring the one-sided nature of the
problem, the parameter $\kappa$ quantifying the price of a unit of shortfall.
\end{remark}

 \section{Illustrative Examples}\label{sec_exam}

We illustrate the framework with two prototypical classes of fixed-income 
instruments: coupon bonds and interest-rate swaps combined with repo financing.

\subsection{Coupon Bonds}

Government coupon bonds are the canonical fixed-income instruments in our setup.
Such a bond is specified by a face value $F>0$ paid at maturity $T$ and coupon
payments $c_1,\dots,c_n$ at dates $0<T_1<\dots<T_n=T$.  
If the bond is indexed by $i$, its cash-flow vector is given by
\begin{equation}\label{eqbondC}
    C_{i,j} = 
    \begin{cases}
        c_n + F, & \text{if } x_j = T, \\
        c_k,     & \text{if } x_j = T_k,\ k=1,\dots,n-1,\\
        0,       & \text{otherwise}.
    \end{cases}
\end{equation}
The corresponding price $P_i$ is observed in the bond market.

\subsection{Interest-Rate Swaps and Repo}

We next illustrate how fixed-income cash flows can be generated by combining
interest-rate swaps with repo financing.\footnote{Repo refers to a standard
repurchase agreement.}

Consider an overnight indexed receiver swap with fixed rate $R$, notional~1, and
maturity $T = n\Delta$, for a fixed accrual period $\Delta > 0$.\footnote{For
formal definitions of overnight indexed swaps (OIS), see, for example,
\cite[Section~2]{fil_tro_13}.}
The holder receives fixed payments $\Delta R$ at dates $T_k = k\Delta \le T$ and 
pays floating interest $\xi_1,\dots,\xi_m$ at dates 
$0<t_1<\dots<t_m = T$, where $\xi_j$ is the repo rate earned over $[t_{j-1},t_j]$.

We construct the following swap--repo strategy.
Investing~1 in the repo market at inception and rolling it forward produces a 
unit payoff at $T$ and floating cash flows $\xi_j$ at the dates $t_j\le T$. Entering the receiver swap at zero initial cost offsets these floating payments 
and yields fixed payments $\Delta R$ at $T_k\le T$. Thus the combined strategy produces exactly the cash flows of a coupon bond with 
face value $F=1$ and coupons $c_k = \Delta R$. Under the absence of dynamic arbitrage, this synthetic bond must have initial 
price~1, since it is generated by a zero-cost swap combined with a unit repo 
investment.\footnote{
This notion of dynamic arbitrage is broader than the static definition in 
Definition~\ref{defNA}.  
If the market price of the bond exceeded~1, one could short the bond and 
replicate it using the swap--repo strategy to lock in a risk-free profit, and 
similarly if the price were below~1.
}
Hence the swap--repo strategy can be encoded in the form \eqref{eqbondC} with 
price $P_i = 1$.

\begin{remark}[Swap spread puzzle]
Under the absence of dynamic arbitrage, the swap--repo replication of coupon
bonds implies that government bonds and interest rate swaps should be priced
using the same discount curve.  
Empirically, however, government bond--swap spreads are persistently nonzero.  
This \emph{swap spread puzzle} is commonly attributed to limits to arbitrage arising
from market frictions, balance-sheet and regulatory constraints, and the safe-asset
premium; see, for example, \cite{chr_mir_21, wu_jar_24, aqu_etal_24}.
\end{remark}

To operationalize the swap--repo strategy in super-replication, suppose there are
$M$ receiver swaps indexed by $i=1,\dots,M$.  
For simplicity we assume that the floating payment dates of the swaps and the
repo roll dates coincide with the fixed-income grid, that is, $x_j=t_j$.
Swap~$i$ has maturity $n_i \Delta$ and fixed rate~$R_i$.

We collect the fixed and floating swap payments in the $M \times N$ matrices
$S$ and $\Xi$, defined by
\[
    S_{i,j} =
    \begin{cases}
        \Delta R_i, & \text{if } x_j = k\Delta,\ k=1,\dots,n_i,\\
        0, & \text{otherwise},
    \end{cases}
    \qquad
    \Xi_{i,j} =
    \begin{cases}
        \xi_j, & \text{if } x_j \le n_i\Delta,\\
        0, & \text{otherwise}.
    \end{cases}
\]
We also define an $M \times N$ face-value matrix~$F$ by
\[
    F_{i,j} =
    \begin{cases}
        1, & \text{if } x_j = n_i\Delta,\\
        0, & \text{otherwise}.
    \end{cases}
\]
The coupon-bond cash-flow matrix corresponding to the swap--repo strategy is
\[
    C = S + F
      = \underbrace{S - \Xi}_{\text{swap net cash flows}}
        + 
        \underbrace{\Xi + F}_{\text{repo net cash flows}},
\]
and the initial prices of these synthetic bonds satisfy $P = \boldsymbol{1}$,
since the swap has zero cost and the repo investment requires one unit of initial
funding.\footnote{
The cash flows $\Xi + F$ may also be interpreted as those of floating-rate notes
with matching payment and maturity dates.
}

\medskip

Suppose that there is no strict arbitrage~\ref{wNA1} and that super-replication is feasible~\ref{F1}, and let $q$ be a minimizer of \eqref{suprepeq}, which exists by Theorem~\ref{thm_dual}. The super-replication price of the liabilities is therefore $q^{\top}\boldsymbol{1}$.
To implement the corresponding swap--repo strategy in practice, proceed as
follows.

\medskip\noindent
\textbf{At inception $x_0 = 0$:}
\begin{itemize}
    \item Invest (lend or borrow) the amount $q^{\top}\boldsymbol{1}$ in the repo market.
    \item Enter $q_i$ units of receiver swap $i$ (zero cost).
\end{itemize}

\noindent
\textbf{At each payment date $x_j$:}
\begin{enumerate}
    \item\label{t1} Receive (or pay) floating interest and notional 
        $\sum_{i: x_j \le n_i\Delta} q_i (\xi_j + 1)$ from the repo market.
    \item\label{t2} Reinvest $\sum_{i: x_j < n_i\Delta} q_i$ in the repo market.
    \item\label{t3} Offset the floating legs of the swaps, contributing 
        $(q^{\top}\Xi)_j = \sum_{i: x_j \le n_i\Delta} q_i \xi_j$.
    \item\label{t4} Receive (or pay) fixed swap interest $(q^{\top}S)_j$.
\end{enumerate}

The repo-related transactions in \ref{t1}--\ref{t2} generate the cash flow
$(q^{\top}(\Xi+F))_j$, while the swap-related transactions in \ref{t3}--\ref{t4} generate
$(q^{\top}(S-\Xi))_j$.
Altogether, the net cash flow is $(q^{\top}(S+F))_j = (q^{\top}C)_j$, which by
construction satisfies
\[
    (q^{\top}C)_j \;\ge\; Z_j,
\]
and thus super-replicates the expected liability cash flows.

\section{A Treasury Cross-Section}\label{sec_num}

This section applies the framework to cross-sections of US Treasury securities.
It asks whether observed price systems satisfy the hypotheses of
Section~\ref{sec_FTFIP}, what follows for the construction of discount curves,
and how tight the replication bounds of Section~\ref{sec_repli} are in practice.

Results are reported in detail for a single quote date, 20~November 2025, and
the whole exercise is then repeated on five further dates to confirm that the
findings are not particular to it. All six are collected in
Table~\ref{tab_lop}.

\subsection{Data}\label{sec_num_data}

We use the CRSP US Treasury Database, daily files. The universe consists of
fully taxable, non-flower, noncallable bills, notes and bonds, excluding TIPS
and floating-rate notes. Following \cite{Filipovic2022} we drop the two most
recently issued securities at each original maturity of 2, 3, 4, 5, 7, 10, 20
and 30 years, so as to remove the on-the-run liquidity premium, and we drop
securities maturing within 90 days. On 20~November 2025 this leaves $M = 349$
instruments---27 bills, 219 notes and 103 bonds---with $N = 249$ distinct
cash-flow dates and a longest maturity of 29.50 years. Prices are full invoice
prices per 100 of face value: end-of-day clean prices, themselves bid--ask
midpoints, plus accrued interest. The entries of $C$ are in the same units.

The cash flows interlock, which is what makes the identification question
nontrivial. Payment dates cluster on the fifteenth and on month ends, so
securities of different original tenors share dates, and adjacent dates are
often a fortnight or less apart. The grid also thins with maturity, from about
two dates a month inside five years to the quarterly
February--May--August--November cycle of the long bonds beyond ten years.

\subsection{Observed prices are not arbitrage-free}\label{sec_num_lop}

By Theorem~\ref{thm_LoP} the law of one price holds if and only if
$P\in\Ima(C)$. It therefore holds for \emph{every} price vector exactly when
$\rank(C)=M$. When $\rank(C)<M$, the price vectors satisfying it form a proper
subspace of $\R^{M}$, hence a Lebesgue-null set.

On these cross-sections $C$ is rank-deficient by a wide margin: on 20~November
2025, $\rank(C)=240$ against $M=349$ and $N=249$, and $\rank(C)<\min(M,N)$ on
all six dates. The reason is structural. The cash-flow row of instrument $i$
consists of a constant coupon paid on a schedule $S_i$ of dates together with
the principal paid at maturity, so the rows of any group of securities sharing
a schedule lie in the two-dimensional span of the coupon strip and the
principal payment, however many such securities there are. Among the 349
instruments there are only 234 distinct schedules, and the schedules are
heavily nested, a bond maturing early paying on a subset of the dates of a
longer one.

Rank deficiency does not by itself imply that a \emph{given} price vector
violates the law of one price: a rank-one cash-flow matrix with proportionally
quoted instruments satisfies it perfectly. It says only that nothing guarantees
it, so the question must be settled by measurement. Table~\ref{tab_lop} settles
it. On every date the observed price vector lies outside $\Ima(C)$, at an
orthogonal distance of 2.0 to 6.4 cents per 100 of face value in
root-mean-square. The law of one price fails, and a fortiori
$\Gcal=\emptyset$, so by Theorem~\ref{thm:wNA} there is a strict arbitrage.

The consequences are not confined to Section~\ref{sec_FTFIP}. Let $q_0$ be a
strict arbitrage and suppose $\Qcal\neq\emptyset$, as is the case here. If
$q\in\Qcal$ then $q+tq_0\in\Qcal$ for every $t\ge 0$, since
$(q+tq_0)^{\top}C\ge q^{\top}C\ge Z$, while
$(q+tq_0)^{\top}P = q^{\top}P + t\,q_0^{\top}P \to -\infty$ as $t\to\infty$.
The linear program \eqref{suprepeq} is therefore unbounded and
$\pi^{+}(Z)=-\infty$. An arbitrage-free price system is not a convenience but a
precondition for Section~\ref{sec_repli}.

\begin{table}[t]
\centering
\tcapfig{The law of one price on six Treasury cross-sections}
\begin{tabular}{lrrrrrcr}
\toprule
& & & & \multicolumn{2}{c}{$\operatorname{dist}(P,\Ima(C))$} & & \\
\cmidrule(lr){5-6}
Date & $M$ & $N$ & $\rank(C)$ & RMSE & max & $P\in\Ima(C)$ & $\min_j\hat v_j$\\
\midrule
2024-09-12 & 343 & 248 & 234 & 0.064 & 0.400 & no & 0.308 \\
2025-03-31 & 346 & 246 & 234 & 0.039 & 0.289 & no & 0.256 \\
2025-06-30 & 347 & 247 & 236 & 0.036 & 0.320 & no & 0.238 \\
2025-09-30 & 351 & 249 & 239 & 0.028 & 0.220 & no & 0.241 \\
2025-11-20 & 349 & 249 & 240 & 0.026 & 0.184 & no & 0.238 \\
2025-12-31 & 352 & 248 & 239 & 0.020 & 0.129 & no & 0.229 \\
\bottomrule
\end{tabular}
\bnotefig{Distances are orthogonal residuals of $P$ on $\Ima(C)$, per 100 of
face value. The last column reports the smallest entry of the estimated
discount vector $\hat v$ of Section~\ref{sec_num_est}; it is strictly positive
on every date, so that the estimated price system satisfies~\ref{sNA3}.}
\label{tab_lop}
\end{table}

It follows that no estimator reproducing the observed prices exactly can exist,
since such an estimator would return an element of $\Gcal$. Bootstrapping in
particular does not apply, and this is a matter of feasibility rather than of
the stability of the resulting forward rates. That the classical procedures
nonetheless work is explained by the same observation: they are applied not to
the full cross-section but to a selected subset of bonds, and the selection is
what restores $\rank(C)=M$. Table~\ref{tab_boot} illustrates this, retaining the
longest bond in each maturity bucket. On a thinned universe the law of one price
holds automatically, whatever the prices, and an exact-fit strictly positive
discount vector exists. The failure documented above is a property of the full
cross-section, not of Treasury prices.

\begin{table}[t]
\centering
\tcapfig{Bootstrapping-style subsets, 20~November 2025}
\begin{tabular}{lrrcc}
\toprule
Bucket width & Bonds $M$ & $\rank(C)$ & Law of one price & $\Gcal\neq\emptyset$ \\
\midrule
1 year   &  30 &  30 & holds for every $P$ & yes \\
6 months &  56 &  56 & holds for every $P$ & yes \\
3 months & 102 & 102 & holds for every $P$ & yes \\
\bottomrule
\end{tabular}
\bnotefig{One bond per maturity bucket, the longest in each. Since
$\rank(C)=M$, the map $v\mapsto Cv$ is onto $\R^{M}$ and \ref{LoP2} holds for
any price vector; the last column reports that a nonnegative solution also
exists at the observed prices.}
\label{tab_boot}
\end{table}

\subsection{Estimated prices}\label{sec_num_est}

Prices must therefore be treated as observed with error, and a discount curve
estimated rather than bootstrapped. We use the kernel ridge estimator of
\cite{Filipovic2022}, which trades pricing error against a smoothness reward
under weights chosen so that the criterion is, to first order, a mean-squared
yield error.\footnote{Kernel parameters as in the replication code accompanying
\cite{Filipovic2022}: $\alpha=0.05$, $\delta=0$, ridge parameter 1, and
duration-based weights. Estimated curves for the US Treasury market are
available at \url{https://www.discount-bond-data.org/}.}

The estimator returns a strictly positive $\hat v$, with smallest entry between
$0.229$ and $0.308$ across the six dates (Table~\ref{tab_lop}). Condition
\ref{sNA3} therefore holds by construction for the estimated price system
$P^\ast\coloneq C\hat v$, hence also no arbitrage~\ref{sNA1} by
Theorem~\ref{thm:NA}, and Assumption~\ref{ass_ND} is satisfied by any bill.
Everything below is computed for $(P^\ast, C)$. The residual $P-P^\ast$, whose
mean absolute value ranges from 3.5 to 7.8 cents across the six dates, is the
pricing error whose cross-sectional dispersion \cite{hu_pan_wang_13} interpret
as a liquidity signal.

Estimation does not remove the multiplicity of Remark~\ref{rem_mult}. Write
$\Gcal^\ast\coloneq\{v\in\R^N_+ : Cv=P^\ast\}$ for the admissible set of the
estimated price system, and let
\[
    \Acal \coloneq \{v\in\R^N : Cv = P^\ast\} = \hat v + \ker(C)
\]
be the affine set it sits in. Since $\hat v>0$, a whole neighbourhood of
$\hat v$ in $\Acal$ lies in $\R^N_{+}$, so $\Gcal^\ast$ has nonempty interior
relative to $\Acal$ and the two have the same affine hull. A coordinate
functional $v\mapsto v_j$ is constant on $\Acal$, and hence on $\Gcal^\ast$,
precisely when it annihilates $\ker(C)$. Accordingly, put
\begin{equation}\label{eqJ_supp}
    J \coloneq \{\, j\le N \;:\; w_j\neq 0 \text{ for some } w\in\ker(C) \,\} ,
\end{equation}
the set of coordinates on which $\ker(C)$ does not vanish identically. This is a
well-defined subset of dates, and it is what distinguishes the discount factors
the prices determine from those they do not; it is \emph{not} a claim that
$\ker(C)$ is spanned by standard basis vectors, and indeed no $e_j$ lies in
$\ker(C)$ here. Equivalently, $J$ indexes the nonzero columns of any matrix
whose rows span $\ker(C)$, and a generic element of $\ker(C)$ is nonzero in
every coordinate of $J$ at once.

On 20~November 2025, $\dim\ker(C)=9$ and $|J|=26$: the prices determine 223 of
the 249 discount factors exactly and leave 26 free. The 26 fall into two groups.
Most occur in pairs about a month apart, at which the same small set of
instruments pays in fixed proportion, so that the market prices the combination
of the two payments but not their split. The rest form a block of quarterly
dates between 9.7 and 13.0 years, where the notes have run off and only the
coupons of the long bonds remain. This is the configuration anticipated in
Remark~\ref{rem_mult}.

The estimator picks $\hat v$ out of $\Gcal^\ast$ by its smoothness reward. The
pricing constraints do not: every element of $\Gcal^\ast$ reproduces $P^\ast$
just as well, and the market, summarised by $P^\ast$, does not distinguish among
them. By Remark~\ref{rem_selection} the bounds computed next therefore range
over all of $\Gcal^\ast$ and not over the selected element, which is what makes
them independent of the estimator.

\subsection{Replication of an annuity}\label{sec_num_repl}

As liability we take a level annuity paying one unit a year for 29 years. Its
payment dates are the traded cash-flow dates nearest to 20~November of each year
from 2026 to 2054, a median of five days from the nominal date, and they lie
between 0.99 and 29.01 years. Placing the payments on traded dates is what makes
super-replication feasible at all: by the necessary condition following
Theorem~\ref{thm_WP}, a payment at a date carrying no instrument cash flow would
give $\pi^{+}(Z)=+\infty$. This is also why the annuity stops at 29 years. A
thirtieth payment in November 2055 would fall beyond 29.50 years, the longest
traded maturity, and could be accommodated only through the cash-flow
aggregation of Remark~\ref{rem_agg}, which reassigns the payments of the longest
bonds to the terminal liability date.

On 20~November 2025,
\[
    \pi^{-}(Z) = 13.3224, \qquad Z\hat v = 15.7893, \qquad \pi^{+}(Z) = 16.4811 .
\]
By Corollary~\ref{cor_interval} the outer two values are sharp: no price outside
the interval can be assigned to the liability without creating a strict
arbitrage, and both endpoints are attained. The middle value is the liability
valued at the estimated curve. The distinction is worth drawing. The
mark-to-model value is what a curve estimate delivers; the interval is what a
static portfolio of the traded instruments can guarantee. The gap between them
measures the part of the valuation that rests on the smoothness reward of the
estimator rather than on traded prices, which for economic capital and for the
regulatory frameworks of Section~\ref{sec_intro} is precisely the quantity of
interest. The mark lies inside the interval on all six dates.

\begin{figure}[t]
\centering
\includegraphics[width=\textwidth]{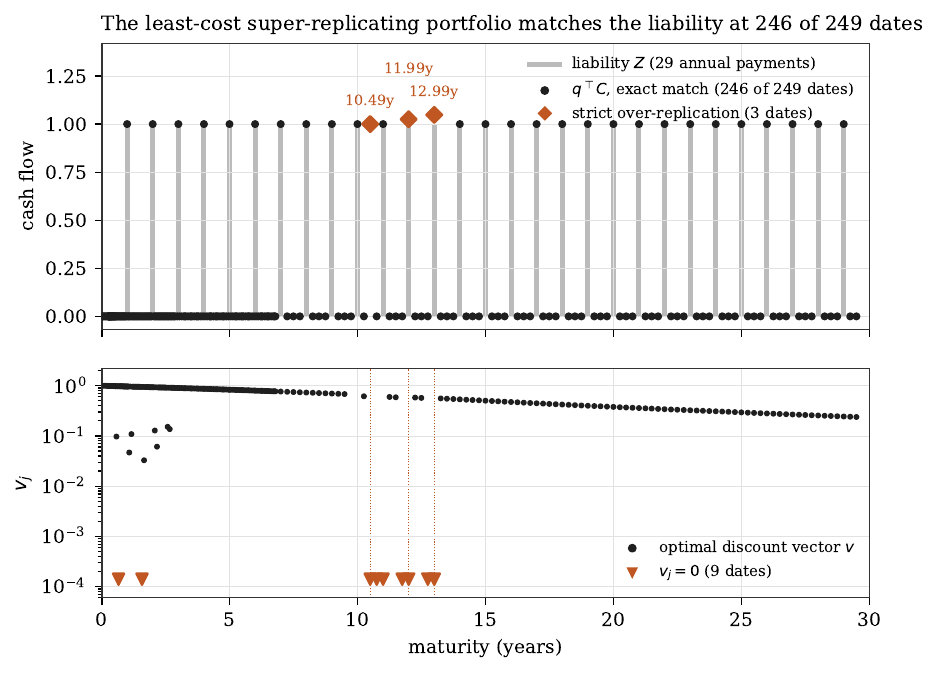}
\tcapfig{Least-cost super-replication of an annuity}
\bnotefig{Upper panel: the liability cash flows $Z$ and the cash flows
$q^{\top}C$ of the least-cost super-replicating portfolio, 20~November 2025.
Lower panel: the optimal discount vector $v$ of the dual problem in
\eqref{eqdual}, on a logarithmic scale, with its nine zero entries marked.
Strict over-replication occurs only where $v_j=0$, as required by
Theorem~\ref{thm_cs}\ref{cs1}.}
\label{fig_repl}
\end{figure}

The optimal portfolio is far tighter than the interval suggests, as
Theorem~\ref{thm_cs} predicts. It matches the liability exactly at 246 of the
249 dates. Strict over-replication occurs at three dates only, at 10.49, 11.99
and 12.99 years, with excesses of $1.000$, $0.025$ and $0.048$ units. All three
lie in the block of dates in $J$, and all three carry $v_j=0$ at the optimal
dual solution, of which there are exactly nine, as
Theorem~\ref{thm_cs}\ref{cs1} requires; the excess cash flows are therefore
costless. Numerically the duality gap in \eqref{eqdual} is $1.1\times 10^{-14}$,
complementary slackness holds to $5.6\times 10^{-15}$, and
$q^{\top}P^\ast=\sum_j Z_j v_j$ to fourteen digits, confirming
Theorem~\ref{thm_cs}\ref{cs2}. Figure~\ref{fig_repl} displays both.

Theorem~\ref{thm_cs}\ref{cs4} guarantees at least $\rank(C)=240$ exactly matched
dates here, against the 246 observed. Note that the guarantee is governed by the
rank and not by the number $M=349$ of instruments: with 234 distinct payment
schedules among them, 109 of the securities are redundant in the sense that
their cash flows are reproduced by the others, and redundant instruments enlarge
neither the set of attainable cash flows nor the number of dates that can be
matched.

Across the six dates the exactly matched share ranges from 95.9\% to 98.8\%,
above the corresponding guarantee $\rank(C)$ in every case. The width of the
interval is less stable, ranging from $0.17$ to $7.54$. This is not noise:
strict over-replication occurs between 9.5 and 13.6 years on every date, and the
width depends on whether the annuity payment dates happen to fall on the
quarterly coupon cycle of the long bonds. Quarter-end quote dates are the least
favourable in this respect: a yearly date falling on 30~June, 30~September or
31~December sits some 46 days from every long-end cash-flow date, almost exactly
midway between two coupons of the long bonds. The width is a joint property of the liability and the
traded universe, as Corollary~\ref{cor_interval} and
Corollary~\ref{cor_collapse} describe.

Finally, the bounds are insensitive to the estimator. Varying the kernel
parameter $\alpha$ over $\{0.05,0.10\}$ and the ridge parameter over
$\{0.1,1,10\}$ moves the root-mean-square pricing error by a factor of more than
two, from 6.2 to 13.4 cents, while $\pi^{+}(Z)-\pi^{-}(Z)$ varies only between
$3.141$ and $3.165$ and $Z\hat v$ by less than a basis point. The fit is a
property of the method; the bounds are a property of the market.

\section{Conclusion and Outlook}\label{sec_conc}

This paper develops a static, model-free framework for fixed-income pricing and
the replication of liabilities.  We establish a fundamental
theorem of fixed-income pricing, characterizing arbitrage-free prices through
the existence of a discount curve, and apply this framework to the
super-replication of expected liability cash flows.

Duality shows that the resulting price bounds are attained and hence cannot be
improved, that they collapse precisely when the liability is attainable, and
that some least-cost super-replicating portfolio matches the liability exactly
at no fewer dates than the rank of the cash-flow matrix. Uniqueness of the least-cost
portfolio holds for almost every arbitrage-free price vector, and a canonical
selection is available in the remaining cases through the asymmetric quadratic
hedging criteria of Section~\ref{sec_asym}.

Section~\ref{sec_num} implements the framework on cross-sections of US Treasury
securities. The cash-flow matrix is rank-deficient, so the price vectors
satisfying the law of one price form a null set, and the observed prices are
found to lie outside it on every date examined: no discount curve reproduces
them exactly, and one must be estimated. Taking the estimated price system as
input, the least-cost super-replicating portfolio of an annuity liability
matches the liability at between 95.9\% and 98.8\% of cash-flow dates, and the
resulting bounds are insensitive to the estimator even where its pricing error
is not.

Several directions remain open. First, the illustration
here uses a stylised annuity; a systematic study across realistic insurance
liability profiles, and across markets in which the traded universe is thinner
than the US Treasury market, would be of practical interest. Second, the static
framework here is by construction a buy-and-hold framework, which matches the
institutional reality of liability-driven investment; relating it to dynamic
hedging strategies under funding and collateral constraints remains an
interesting question.

Overall, the results provide a unified static foundation for fixed-income
pricing and liability replication, with implications for asset--liability
management, discount-curve construction, and economic and regulatory capital
assessment.

\section*{Declarations}

\paragraph{Data availability.}
The findings of Section~\ref{sec_num} rest on the CRSP US Treasury Database.
These data were used under licence and cannot be redistributed by the author;
they are available from the Center for Research in Security Prices to holders of
a subscription. Estimated discount curves for the US Treasury market are
publicly available at \url{https://www.discount-bond-data.org/}. The code that
reproduces every quantity and the figure reported in Section~\ref{sec_num} is
available from the author on request.

\paragraph{Funding.}
The author received no specific funding for this work.

\paragraph{Conflict of interest.}
The author declares no conflict of interest.

\paragraph{Ethics.}
This research did not involve human participants or animals, and no ethical
approval was required.

\paragraph{Generative AI.}
In preparing this paper the author used Claude Opus 5 (Anthropic) to write and
run the code producing the numerical results and figure of
Section~\ref{sec_num}, and to draft parts of the text of that section. No other
part of the paper was generated by these tools. The author checked all reported
results and takes full responsibility for the content of the paper.

\begin{appendix}
\section{Proofs}
This appendix collects all proofs.

\subsection{Proof of Theorem~\ref{thm_LoP}}
The law of one price is equivalent to $\ker (C^\top)\subseteq\ker (P^\top)$, which again is equivalent to $P\in\Ima (C)$.

\ref{LoP1}$\Rightarrow$\ref{LoP2}: By the above there exists a $v\in\R^N$ such that $P=Cv$. Given such a $v$, we obtain a curve $g:[0,\infty)\to \R$ with $g(0)=1$ by linearly interpolating $g(\bm x)=v$ and $g(0)=1$, for $x\in [0,x_N]$, and setting $g(x)=\e^{-y_\infty(x-x_N)}g(x_N)$ for $x>x_N$, for some auxiliary long maturity yield $y_\infty\ge 0$.

\ref{LoP2}$\Rightarrow$\ref{LoP1}: This follows trivially by the above.

\subsection{Proof of Theorem \ref{thm:wNA}}
\ref{wNA1}$\Leftrightarrow$\ref{wNA2}: this is elementary.

\ref{wNA2}$\Leftrightarrow$\ref{wNA3}: Farkas' Lemma \cite[Corollary 22.3.1]{roc_97} states that \ref{wNA2} holds if and only if there exists a vector $v\ge 0$ such that $P=C v$. Given such a $v$, we obtain a desired nonnegative curve $g:[0,\infty)\to [0,\infty)$ by linearly inter-extrapolating $g$ as in the proof of Theorem \ref{thm_LoP}.

For the last statement, let $q$ be such that $q^\top C=0$. Now apply no strict arbitrage to $q$ and $-q$, which implies $q^\top P\ge 0$ and $-q^\top P\ge 0$, and thus $q^\top P=0$.

\subsection{Proof of Theorem \ref{thm:NA}}
We first record two elementary facts.

\begin{lemma}\label{lem_half}
Let $a,P\in\R^M$ be such that $q^\top a\le 0$ for all $q\in\R^M$ with
$q^\top P\le 0$. Then $a = sP$ for some $s\ge 0$.
\end{lemma}

\begin{proof}
If $P=0$, the hypothesis gives $q^\top a\le0$ for all $q\in\R^M$, hence $a=0$
and we may take $s=0$. Let $P\neq0$. For any $q$ with $q^\top P=0$ we also have
$(-q)^\top P=0$, so that $q^\top a\le0$ and $-q^\top a\le0$, whence
$q^\top a=0$. Thus $a\in(P^\perp)^\perp=\spn\{P\}$, say $a=sP$. Taking $q=-P$
gives $q^\top P=-\|P\|^2\le0$ and therefore $-s\|P\|^2=q^\top a\le0$, so
$s\ge0$.
\end{proof}

\begin{lemma}\label{lem_S1W1}
Under Assumption~\ref{ass_ND}, \ref{sNA1} implies \ref{wNA1}.
\end{lemma}

\begin{proof}
Assume \ref{sNA1} and suppose $q_0$ is a strict arbitrage, $q_0^\top P<0$ and
$q_0^\top C\ge0$. If $q_0^\top C\neq0$, then $q_0$ is an arbitrage,
contradicting \ref{sNA1}. Hence $q_0^\top C=0$. Let $p$ be as in
Assumption~\ref{ass_ND}, so that $p^\top C\ge0$ and $p^\top C\neq0$, and set
$q_t\coloneq p+t\,q_0$ for $t\ge0$. Then $q_t^\top C=p^\top C$ is nonnegative
and nonzero for every $t\ge 0$, while $q_t^\top P=p^\top P+t\,q_0^\top P\to-\infty$
as $t\to\infty$. For $t$ large enough, $q_t$ is thus an arbitrage, which
contradicts \ref{sNA1}.
\end{proof}

\ref{sNA1}$\Leftrightarrow$\ref{sNA2}: this is elementary, and uses neither
Assumption~\ref{ass_ND} nor any separation argument.

\ref{sNA3}$\Rightarrow$\ref{sNA2}: Let $g>0$ be a positive discount curve such that $P=C g(\bm x)$. Let $w=C^\top q\in \overline\Ccal\cap \R^N_+$. Then $0\ge q^\top P = q^\top C g(\bm x)\ge 0$, as also $q^\top C\ge 0$. This implies $q^\top C =0$, which proves \ref{sNA2}. This implication, too, holds without Assumption~\ref{ass_ND}.

\ref{sNA2}$\Rightarrow$\ref{sNA3}: as $\R^N_+$ is a pointed convex cone, i.e., $\R^N_+\cap (-\R^N_+)=\{0\}$, Klee's separation theorem \cite[Theorem (2.5)]{kle_55} applies, stating that there exists a $v\in\R^N$ such that $v^\top w>0$ for all $w\in\R^N_+\setminus\{0\}$ and $z^\top v\le 0$ for all $z\in\overline\Ccal$. From the first it follows that $v_j>0$ for all $j\le N$. From the second it follows that $q^\top Cv\le 0$ for all $q\in\R^M$ with $q^\top P\le 0$, and therefore $Cv = sP$ for some $s\ge 0$, by Lemma~\ref{lem_half}.

If $s>0$, we set $g(\bm x)=v/s$, which is strictly positive and satisfies $C\,g(\bm x)=P$. If $s=0$, then $Cv=0$. By \ref{sNA2}$\Leftrightarrow$\ref{sNA1}, Lemma~\ref{lem_S1W1} and Theorem~\ref{thm:wNA} there exists $v_0\in\R^N_+$ with $Cv_0=P$, and we set $g(\bm x)=v_0+v$, which is strictly positive and satisfies $C(v_0+v)=P$. In either case, inter-extrapolating $g$ as in the proof of Theorem \ref{thm:wNA} gives the desired positive discount curve $g:[0,\infty)\to (0,\infty)$.

The last statement follows from \ref{sNA1}$\Leftrightarrow$\ref{sNA3}, the trivial implication \ref{sNA3}$\Rightarrow$\ref{wNA3}, and Theorem~\ref{thm:wNA}. Example~\ref{ex_ND} shows that Assumption~\ref{ass_ND} cannot be omitted.

\subsection{Proof of Theorem~\ref{thm_WP}}
\ref{F1}$\Leftrightarrow$\ref{F2}: this follows from Farkas' lemma (in the form of \cite[Theorem 22.1]{roc_97}) applied to the system $C^\top q\ge Z^\top$.

The last statement follows by setting $q= t e_i$, where $e_i$ is the $i$th standard basis vector in $\R^M$ and $t=\frac{\max\{\max_j Z_j,0\}}{\min\{C_{i,j}\,:\,C_{i,j}>0\}}\ge 0$. Indeed, for $j$ with $C_{i,j}>0$ we have $t\,C_{i,j}\ge \max\{\max_k Z_k,0\}\ge Z_j$, and for $j$ with $C_{i,j}=0$ we have $Z_j=0=t\,C_{i,j}$ by assumption.

\subsection{Proof of Theorem~\ref{thm_dual}}
Problem \eqref{suprepeq} is the linear program $\min\{P^\top q:\ C^\top q\ge Z^\top,\ q\in\R^M\}$, whose dual is $\max\{Z v:\ Cv=P,\ v\ge 0\}$, that is, $\max\{Zv:\ v\in\Gcal\}$. By \ref{wNA1} and Theorem~\ref{thm:wNA} the dual is feasible, $\Gcal\neq\emptyset$; in particular $\sup_{v\in\Gcal}Zv>-\infty$. The duality theorem of linear programming then gives the first identity in \eqref{eqdual}: if the primal is infeasible, the dual is unbounded and both sides equal $+\infty$; if the primal is feasible, that is, if \ref{F1} holds, then both values are finite, equal, and attained.

For the second identity, note that $q$ is feasible for \eqref{subrepeq} if and only if $-q$ is feasible for the super-replication problem associated with the liability $-Z$, and $q^\top P = -(-q)^\top P$; hence $\pi^-(Z)=-\pi^+(-Z)$. Applying the first identity to the liability $-Z$ gives $\pi^-(Z) = -\sup_{v\in\Gcal}(-Z)v=\inf_{v\in\Gcal}Zv$, together with the corresponding finiteness and attainment statement. Finally, $\inf_{v\in\Gcal}Zv\le\sup_{v\in\Gcal}Zv$ because $\Gcal\neq\emptyset$.

\subsection{An auxiliary lemma}
The following lemma isolates the argument underlying the compactness statements used below.

\begin{lemma}\label{lem_rec}
Assume there is no arbitrage~\ref{sNA1} and $\ker(C^\top)=\{0\}$. Then $u^\top P>0$ for every $u\in\R^M\setminus\{0\}$ with $u^\top C\ge 0$. Consequently, for every $c\in\R$ the set $\Qcal\cap\{q:\ q^\top P\le c\}$ is compact.
\end{lemma}

\begin{proof}
Let $u\neq 0$ with $u^\top C\ge 0$. Since $\ker(C^\top)=\{0\}$ we have $u^\top C\neq 0$. If $u^\top P\le 0$, then $u$ would be an arbitrage, contradicting \ref{sNA1}; hence $u^\top P>0$.

The set $\Qcal_c\coloneq\Qcal\cap\{q:\ q^\top P\le c\}$ is closed and convex, with recession cone $\rec(\Qcal_c)=\{u:\ u^\top C\ge 0,\ u^\top P\le 0\}=\{0\}$ by the above. A closed convex set whose recession cone is trivial is bounded, see \cite[Theorem 8.4]{roc_97}, and hence $\Qcal_c$ is compact.
\end{proof}

\subsection{Proof of Corollary~\ref{cor_interval}}
Enlarge the market by an instrument with cash-flow row $Z$ and price $p$, that is, consider $\tilde C=\begin{bmatrix}C\\ Z\end{bmatrix}$ and $\tilde P=\begin{bmatrix}P\\ p\end{bmatrix}$. By Theorem~\ref{thm:wNA}, the enlarged market admits no strict arbitrage if and only if there exists $v\in\R^N_+$ with $\tilde C v=\tilde P$, that is, if and only if $v\in\Gcal$ and $Zv=p$. The set of such $p$ therefore equals $\{Zv: v\in\Gcal\}$, which is the image of the nonempty convex set $\Gcal$ under a linear map and hence an interval, whose endpoints are $\pi^{\pm}(Z)$ and are attained whenever finite, by Theorem~\ref{thm_dual}.

\subsection{Proof of Corollary~\ref{cor_collapse}}
If $Z^\top=C^\top q$ for some $q\in\R^M$, then $Zv=q^\top Cv=q^\top P$ for every $v\in\Gcal$, so that $v\mapsto Zv$ is constant on $\Gcal$ and $\pi^-(Z)=\pi^+(Z)=q^\top P$ by Theorem~\ref{thm_dual}.

Conversely, assume $\pi^-(Z)=\pi^+(Z)$, so that $v\mapsto Zv$ is constant on $\Gcal$. By \ref{sNA1} and Theorem~\ref{thm:NA} there exists $v^\ast\in\Gcal$ with $v^\ast>0$. For every $w\in\ker(C)$ and every $\epsilon$ small enough in absolute value we have $v^\ast+\epsilon w\in\R^N_+$ and $C(v^\ast+\epsilon w)=P$, that is, $v^\ast+\epsilon w\in\Gcal$, and therefore $Zw=0$. Hence $Z^\top\in\ker(C)^\perp=\Ima(C^\top)$.

\subsection{Proof of Theorem~\ref{thm_cs}}
\ref{cs1} and \ref{cs2}: complementary slackness for the primal--dual pair in the proof of Theorem~\ref{thm_dual} gives $v_j[(q^\top C)_j-Z_j]=0$ for all $j\le N$. Consequently $q^\top P=q^\top C v=\sum_{j\le N}(q^\top C)_j v_j=\sum_{j\le N} Z_j v_j$.

\ref{cs3}: if $v>0$, then \ref{cs1} forces $(q^\top C)_j=Z_j$ for all $j\le N$.

\ref{cs4}: by Theorem~\ref{thm_dual} the linear program \eqref{suprepeq} is feasible with finite value, and its solution set $\Qcal^\ast$ is nonempty. For $q\in\Qcal^\ast$ write $I(q)\coloneq\{j\le N:\ (q^\top C)_j=Z_j\}$ for the active set and $V(q)\coloneq\spn\{C_{:,j}:\ j\in I(q)\}\subseteq\Ima(C)$. Since there are finitely many active sets, we may choose $q\in\Qcal^\ast$ with $\dim V(q)$ maximal. We claim $V(q)=\Ima(C)$.

Suppose not, and pick $u\in\Ima(C)\cap V(q)^\perp$ with $u\neq 0$. Then $u^\top C_{:,j}=0$ for every $j\in I(q)$. Moreover $u^\top C\neq 0$: otherwise $u\in\ker(C^\top)\cap\Ima(C)=\{0\}$, as $\Ima(C)=\ker(C^\top)^\perp$. For $|t|$ small the point $q+tu$ lies in $\Qcal$, because the constraints in $I(q)$ remain active and those outside $I(q)$ are strict at $q$; since $q$ is optimal and both signs of $t$ are admissible, $u^\top P=0$, and $q+tu\in\Qcal^\ast$.

Let $\bar t\coloneq\sup\{t\ge 0:\ q+tu\in\Qcal\}$. If $\bar t=\infty$ then $(u^\top C)_j\ge 0$ for all $j$, so $u^\top C\ge 0$ and $u^\top C\neq 0$ while $u^\top P=0$; thus $u$ is an arbitrage, contradicting \ref{sNA1}. Hence $\bar t<\infty$, the point $q^\prime\coloneq q+\bar t u$ belongs to $\Qcal^\ast$, and some constraint $j_0\notin I(q)$ is active at $q^\prime$, with $(u^\top C)_{j_0}<0$. In particular $C_{:,j_0}\notin V(q)$, because $u\perp V(q)$. Therefore $I(q)\cup\{j_0\}\subseteq I(q^\prime)$ and $\dim V(q^\prime)>\dim V(q)$, contradicting the choice of $q$.

Hence $V(q)=\Ima(C)$, so $|I(q)|\ge\dim\Ima(C)=\rank(C)$ and at most $N-\rank(C)$ of the inequalities $(q^\top C)_j\ge Z_j$ are strict. If $\ker(C^\top)=\{0\}$, the lineality space $\{u\in\R^M:\ u^\top C=0\}$ of $\Qcal$ is trivial, so $\Qcal$ is pointed; the active gradients at $q$ then span $\R^M$, which is exactly the condition for $q$ to be a vertex of $\Qcal$.

\subsection{Proof of Proposition~\ref{prop_uniq}}
Since $\ker(C^\top)=\{0\}$, the polyhedron $\Qcal$ is pointed and has a finite, nonempty set of vertices $q^1,\dots,q^K$, which depends on $C$ and $Z$ only. Put
\[
    \Ncal\coloneq\bigcup_{1\le k<l\le K}\{\,P\in\R^M:\ (q^k-q^l)^\top P=0\,\},
\]
a finite union of hyperplanes, since $q^k\neq q^l$ for $k\neq l$.

Let $P\in\Pcal_{\rm na}\setminus\Ncal$. By Lemma~\ref{lem_S1W1} and Theorem~\ref{thm_dual}, the solution set $\Qcal^\ast$ of \eqref{suprepeq} is a nonempty face of $\Qcal$. Its vertices are vertices of $\Qcal$ and all have the same price; since $P\notin\Ncal$, there can be only one, say $q^k$. A pointed polyhedron with a single vertex equals that vertex plus its recession cone, so that $\Qcal^\ast=q^k+\rec(\Qcal^\ast)$. If there were $u\in\rec(\Qcal^\ast)\setminus\{0\}$, then $u^\top C\ge 0$ and $u^\top P=0$, contradicting Lemma~\ref{lem_rec}. Hence $\Qcal^\ast=\{q^k\}$.

\subsection{Proof of Proposition~\ref{prop_qexists}}
The map $u\mapsto (u_+)^2$ is convex, nondecreasing and continuously differentiable with derivative $2u_+$, so the middle term in \eqref{eqJ} is convex and continuously differentiable, being a composition with affine maps. Adding the linear term and the strictly convex, coercive term $\lambda\|q\|_2^2$ yields a strictly convex, coercive and continuously differentiable function, which therefore admits a unique minimizer, characterized by $\nabla J_{\kappa,\lambda}=0$. Differentiating,
\[
    \nabla J_{\kappa,\lambda}(q)= P-2\kappa\,C\Bigl(\bigl(Z-q^\top C\bigr)_+\Bigr)^{\!\top}+2\lambda q,
\]
which vanishes exactly when \eqref{eqFOC} holds.

\subsection{Proof of Theorem~\ref{thm_limit}}
(i) Fix $\lambda>0$ and write $\phi(q)\coloneq\sum_{j\le N}\bigl((Z_j-(q^\top C)_j)_+\bigr)^2\ge 0$, so that $\phi(q)=0$ if and only if $q\in\Qcal$, and $F(q)\coloneq q^\top P+\lambda\|q\|_2^2$. The function $F$ is strictly convex and coercive and $\Qcal$ is nonempty, closed and convex by \ref{F1}, so that $q_\lambda=\argmin_{q\in\Qcal}F(q)$ is well defined. Since $J_{\kappa,\lambda}(q_{\kappa,\lambda})\le J_{\kappa,\lambda}(q_\lambda)=F(q_\lambda)\eqqcolon c$ and $J_{\kappa,\lambda}\ge F$, coercivity of $F$ implies that the family $(q_{\kappa,\lambda})_{\kappa>0}$ is bounded, and $\kappa\,\phi(q_{\kappa,\lambda})\le c-\inf F<\infty$ implies $\phi(q_{\kappa,\lambda})\to 0$ as $\kappa\to\infty$. Let $\bar q$ be a limit point along some sequence $\kappa_n\to\infty$. Then $\phi(\bar q)=0$, that is, $\bar q\in\Qcal$, and $F(\bar q)\le\liminf_n J_{\kappa_n,\lambda}(q_{\kappa_n,\lambda})\le c=F(q_\lambda)$. By uniqueness of $q_\lambda$ we obtain $\bar q=q_\lambda$, and boundedness gives $q_{\kappa,\lambda}\to q_\lambda$ as $\kappa\to\infty$.

(ii) By Theorem~\ref{thm_dual} the solution set $\Qcal^\ast$ of \eqref{suprepeq} is nonempty, closed and convex, and it is compact by Lemma~\ref{lem_rec}, being of the form $\Qcal\cap\{q:q^\top P\le\pi^+(Z)\}$. Let $q^\star$ denote its unique minimal-norm element and write $\pi^+=\pi^+(Z)$. For $\lambda>0$,
\[
    q_\lambda^\top P+\lambda\|q_\lambda\|_2^2\le (q^\star)^\top P+\lambda\|q^\star\|_2^2=\pi^++\lambda\|q^\star\|_2^2,
\]
while $q_\lambda^\top P\ge\pi^+$. Hence $\|q_\lambda\|_2\le\|q^\star\|_2$ and $q_\lambda^\top P\le\pi^++\lambda\|q^\star\|_2^2$. The family $(q_\lambda)_{\lambda>0}$ is therefore bounded, and every limit point as $\lambda\downarrow 0$ belongs to $\Qcal^\ast$ and has norm at most $\|q^\star\|_2$, hence equals $q^\star$ by uniqueness of the minimal-norm element. Thus $q_\lambda\to q^\star$ as $\lambda\downarrow 0$.

\end{appendix}

\bibliographystyle{alpha}
\bibliography{bibliography}

@article{CamenzindFilipovic2024,
  title = {Stripping the {S}wiss discount curve using kernel ridge regression},
  volume = {14},
  ISSN = {2190-9741},
  url = {http://dx.doi.org/10.1007/s13385-024-00386-4},
  DOI = {10.1007/s13385-024-00386-4},
  number = {2},
  journal = {European Actuarial Journal},
  publisher = {Springer Science and Business Media LLC},
  author = {Camenzind,  Nicolas and Filipovic,  Damir},
  year = {2024},
  month = jun,
  pages = {371–410}
}

@article{Filipovic2022,
  author    = {Damir Filipovic and Markus Pelger and Ye Ye},
  title     = {Stripping the Discount Curve---a Robust Machine Learning Approach},
  journal   = {Management Science},
  year      = {2025},
  volume    = {0},
  number    = {0},
  doi       = {10.1287/mnsc.2023.01401},
  url       = {https://doi.org/10.1287/mnsc.2023.01401},
  note      = {Articles in Advance}
}

@ARTICLE{aqu_etal_24,
  author =       {Matteo Aquilina and Andreas Schrimpf and Vladyslav Sushko and Dora Xia},
  title =        {Negative interest rate swap spreads signal pressure in government debt absorption},
  journal =      {BIS Quarterly Review},
  year =         {2024},
  month =        {December},
  note =         {\url{https://www.bis.org/publ/qtrpdf/r_qt2412y.htm}},
  doi =          {https://www.bis.org/publ/qtrpdf/r_qt2412y.htm},
}

@ARTICLE{chr_mir_21,
  author =       {Jens H.E. Christensen and Nikola Mirkov},
  title =        {The safety premium of safe assets},
  journal =      {SNB Working Paper 2021--02},
  year =         {2021},
  note =         {\url{https://www.snb.ch/en/publications/research/working-papers/2021/working_paper_2021_02}},
  doi =          {https://www.snb.ch/en/publications/research/working-papers/2021/working_paper_2021_02},
}

@book{coc_01,
  address = {Princeton [u.a.]},
  author = {Cochrane, {John Howland}},
  isbn = {0691074984},
  pagetotal = {XVII, 530},
  ppn_gvk = {322224764},
  publisher = {Princeton Univ. Press},
  title = {Asset pricing},
  url = {http://gso.gbv.de/DB=2.1/CMD?ACT=SRCHA&SRT=YOP&IKT=1016&TRM=ppn+322224764&sourceid=fbw_bibsonomy},
  year = 2001
}

@article {dal_mor_wil_90,
    AUTHOR = {Dalang, Robert C. and Morton, Andrew and Willinger, Walter},
     TITLE = {Equivalent martingale measures and no-arbitrage in stochastic
              securities market models},
   JOURNAL = {Stochastics Stochastics Rep.},
  FJOURNAL = {Stochastics and Stochastics Reports},
    VOLUME = {29},
      YEAR = {1990},
    NUMBER = {2},
     PAGES = {185--201},
      ISSN = {1045-1129},
   MRCLASS = {90A60 (60G99)},
  MRNUMBER = {1041035},
MRREVIEWER = {Nicola\ Mattoscio},
       DOI = {10.1080/17442509008833613},
       URL = {https://doi.org/10.1080/17442509008833613},
}

@article{fil_tro_13,
title = {The term structure of interbank risk},
journal = {Journal of Financial Economics},
volume = {109},
number = {3},
pages = {707-733},
year = {2013},
issn = {0304-405X},
doi = {https://doi.org/10.1016/j.jfineco.2013.03.014},
url = {https://www.sciencedirect.com/science/article/pii/S0304405X13000949},
author = {Damir Filipovic and Anders B. Trolle}
}

@article{filipovic:exact_term_structure,
      author="Filipovic Damir and Willems Sander",
      title="Exact Smooth Term-Structure Estimation",
      journal="SIAM Journal on Financial Mathematics",
      year="2018"
}

@book {foe_sch_16,
    AUTHOR = {F\"ollmer, Hans and Schied, Alexander},
     TITLE = {Stochastic finance},
    SERIES = {De Gruyter Graduate},
   EDITION = {extended},
      NOTE = {An introduction in discrete time},
 PUBLISHER = {De Gruyter, Berlin},
      YEAR = {2016},
     PAGES = {xii+596},
      ISBN = {978-3-11-046345-3; 978-3-11-046344-6; 978-3-11-046346-0},
   MRCLASS = {91-01 (60G42 60H10 60H30 91B30 91G10 91G20)},
  MRNUMBER = {3859905},
       DOI = {10.1515/9783110463453},
       URL = {https://doi.org/10.1515/9783110463453},
}

@article{gur_sac_wri_07,
	author = {G{\"{u}}rkaynak, Refet S. and Sack, Brian and Wright, Jonathan H.},
	doi = {10.1016/j.jmoneco.2007.06.029},
	isbn = {1556-5068},
	issn = {03043932},
	journal = {Journal of Monetary Economics},
	number = {8},
	pages = {2291--2304},
	title = {{The U.S. Treasury yield curve: 1961 to the present}},
	volume = {54},
	year = {2007}}

@article{har_kre_79,
title = {Martingales and arbitrage in multiperiod securities markets},
journal = {Journal of Economic Theory},
volume = {20},
number = {3},
pages = {381-408},
year = {1979},
issn = {0022-0531},
doi = {https://doi.org/10.1016/0022-0531(79)90043-7},
url = {https://www.sciencedirect.com/science/article/pii/0022053179900437},
author = {J.Michael Harrison and David M Kreps}
}

@article {kle_55,
    AUTHOR = {Klee, Jr., V. L.},
     TITLE = {Separation properties of convex cones},
   JOURNAL = {Proc. Amer. Math. Soc.},
  FJOURNAL = {Proceedings of the American Mathematical Society},
    VOLUME = {6},
      YEAR = {1955},
     PAGES = {313--318},
      ISSN = {0002-9939,1088-6826},
   MRCLASS = {46.0X},
  MRNUMBER = {68113},
MRREVIEWER = {M.\ M.\ Day},
       DOI = {10.2307/2032366},
       URL = {https://doi.org/10.2307/2032366},
}

@article{liu_wu_21,
	author = {Yan Liu and Jing Cynthia Wu},
	doi = {https://doi.org/10.1016/j.jfineco.2021.05.059},
	issn = {0304-405X},
	journal = {Journal of Financial Economics},
	number = {3},
	pages = {1395-1425},
	title = {Reconstructing the yield curve},
	url = {https://www.sciencedirect.com/science/article/pii/S0304405X21002531},
	volume = {142},
	year = {2021}}

@article{nel_sie_87,
	author = {Nelson, Charles R and Siegel, Andrew F},
	journal = {Journal of Business},
	number = {4},
	pages = {473--489},
	title = {{Parsimonious Modeling of Yield Curves}},
	volume = {60},
	year = {1987}}

@book {roc_97,
    AUTHOR = {Rockafellar, R. Tyrrell},
     TITLE = {Convex analysis},
    SERIES = {Princeton Landmarks in Mathematics},
      NOTE = {Reprint of the 1970 original,
              Princeton Paperbacks},
 PUBLISHER = {Princeton University Press, Princeton, NJ},
      YEAR = {1997},
     PAGES = {xviii+451},
      ISBN = {0-691-01586-4},
   MRCLASS = {49-02 (26-02 46-02 58-02 90-02)},
  MRNUMBER = {1451876},
}

@techreport{sve_94,
	address = {Cambridge, MA},
	author = {Svensson, Lars E.O.},
	doi = {10.3386/w4871},
	institution = {National Bureau of Economic Research},
	month = {sep},
	title = {{Estimating and Interpreting Forward Interest Rates: Sweden 1992 - 1994}},
	url = {http://www.nber.org/papers/w4871.pdf},
	year = {1994}}

@misc{wu_jar_24,
  author= "Wu, David and Jarrow, Robert A.",
  note   = "Available at SSRN: \url{https://ssrn.com/abstract=4904777}",
  title     = "The {T}reasury--{SOFR} Swap Spread Puzzle Explained",
  year      = "2024",
  doi       = "https://ssrn.com/abstract=4904777",
}

@article{hu_pan_wang_13,
  author  = {Hu, Grace Xing and Pan, Jun and Wang, Jiang},
  title   = {Noise as Information for Illiquidity},
  journal = {The Journal of Finance},
  volume  = {68},
  number  = {6},
  pages   = {2341--2382},
  year    = {2013},
  doi     = {10.1111/jofi.12083},
}
\end{document}